\documentclass[prd, preprint, longbibliography, 12pt]{revtex4-1}

\usepackage{amsmath}
\usepackage{amssymb}
\usepackage{setspace}
\usepackage{graphicx}
\usepackage{natbib}
\usepackage{float}
\usepackage{tensor}
\usepackage[utf8]{inputenc}
\usepackage{amsfonts}
\usepackage{braket}
\usepackage{esint}
\usepackage{breqn}
\usepackage{IEEEtrantools}
\usepackage{xcolor}
\usepackage{ulem}

\begin{document}
 
 %

\begin{center}
{ \large \bf Dark energy as a large scale  quantum gravitational phenomenon
 }


\vskip 0.2 in

{\large{\bf Tejinder P.  Singh }}

\medskip

{\it Tata Institute of Fundamental Research,}
{\it Homi Bhabha Road, Mumbai 400005, India}\\
\bigskip
 {\tt tpsingh@tifr.res.in}

\end{center}

\centerline{\bf ABSTRACT}
\noindent  In our recently proposed  quantum theory of gravity, the universe is made of `atoms' of space-time-matter (STM). Planck scale foam is composed of STM atoms with Planck length as their associated Compton wave-length. The quantum dispersion and accompanying spontaneous localisation of these STM atoms amounts to a cancellation of the enormous curvature on the Planck length scale. However, an effective dark energy term  arises in Einstein equations, of the order required by current observations on cosmological  scales. This happens if we propose an extremely light particle having a mass of about $10^{-33} \ {\rm eV/c^2}$, forty-two orders of magnitude lighter than the proton.  The holographic principle suggests there are about $10^{122}$ such particles in the observed universe. Their net effect on space-time geometry is equivalent to dark energy, this being a low energy quantum gravitational phenomenon. In this sense, the observed dark energy constitutes evidence for quantum gravity. We then invoke Dirac's large number hypothesis to also propose a dark matter candidate having a mass halfway (on the logarithmic scale) between the proton and the dark energy particle, i.e. about $10^{-12}\ {\rm eV/c^2}$.
\bigskip

\bigskip

\bigskip

There ought to exist an equivalent formulation of quantum (field) theory which does not refer to a background classical space-time. This is because an operational definition of the background requires classical matter fields, and these latter are themselves a limiting case of quantum fields \cite{Singh:2006,Singh:2012,Lochan-Singh:2011,Lochan:2012}. The development of such a formulation leads us to a quantum theory of gravity \cite{maithresh2019}. 

The fundamental theory is a non-commutative matrix dynamics of Grassmann matrices, the so-called `atoms of space-time-matter [STM]' which do not make a distinction between space-time and matter. These matrices evolve in the Hilbert space according to a time parameter characteristic of non-commutative geometry, which derives from the Tomita-Takesaki theory \cite{Tomita1, Connes2000, Takesaki1, Takesaki2} and the Cocycle Radon-Nykodym theorem \cite{Nykodym, connes1994}. The STM atoms are described by a Lagrangian dynamics resulting from a well-defined action principle, and they interact via entanglement, and `collisions' \cite{SinghqgV2019}. The dynamics possesses a unique conserved charge known as the Adler-Millard charge \cite{Adler:04}, responsible for the emergence of quantum theory, as follows \cite{maithresh2019}.

The above theory is assumed to operate at the Planck scale - there is no space-time, but one can define a Planck scale foam of space-time-matter. If one is not observing dynamics at the Planck scale, a mean-field dynamics at lower energies is arrived at. This is done by averaging over time-scales much larger than Planck time, using the standard techniques of statistical thermodynamics. This mean field dynamics falls into two classes.

If a length scale associated with the STM atoms is  larger than Planck length, one gets a quantum theory of gravity for the bosonic (gravity) and fermionic (matter) aspects of the STM atoms. These degrees of freedom evolve with respect to the characteristic  time parameter of non-commutative geometry. Quantum gravity is not exclusively a Planck scale phenomenon, but relevant even at much lower energies if the (quantum) gravity associated with an STM atom cannot be neglected
\cite{SinghqgV2019}. In particular this can happen, as is the case in this paper, if the Compton wavelength associated with the STM atom is of the order of the size of the observed universe.

In the other extreme limit, the entanglement of a very large number of STM atoms results in a rapid `spontaneous localisation', giving rise to a classical space-time geometry driven by point matter sources, and obeying the laws of classical general relativity. Ordinary space-time is recovered, but the non-commutative time parameter is lost in the classical limit.

Consider the STM atoms which have not undergone localisation. Standard quantum field theory is recovered by taking their matter degrees of freedom from quantum gravity, ignoring their gravity, and taking space-time from the above classical limit of other STM atoms.

In the framework of this theory, we propose that dark energy, which causes acceleration of the expanding universe,  consists of about $10^{122}$ STM atoms, each having an associated mass of $10^{-33}\ {\rm eV/c^2}$. Hence the  Compton wavelength $\hbar/mc$ of such an STM atom is of the order of $10^{28}\ {\rm cm}$, comparable to the size of the observed universe. An STM atom is nothing but an elementary particle whose gravitational aspect cannot be distinguished from the particle aspect [hence space-time-matter]. An atom labelled by the matrix/operator variable $q=q_B + q_F$, with $q_B$ and $q_F$ being its bosonic and fermionic parts, is described by the action principle \cite{maithresh2019}
 \begin{equation}
\frac{L_p}{c} \frac{S}{C_0}  =  \frac{1}{2} \int d\tau \; Tr \bigg[\frac{L_p^2}{L^2c^2}\; (\dot{q}_B +\beta_1 \dot{q}_F)\;(\dot{q}_B +\beta_2 \dot{q}_F) \bigg]
\label{funacn}
\end{equation}
The $\beta$ matrices are constant matrices, proportional to $L_p^2/L^2$, where $L$ is a fundamental length associated with the STM atom, subsequently identified with Compton wavelength. For a detailed description the reader is referred to \cite{maithresh2019}. For this action, the first integrals of the equations of motion, with evolution with respect  to the Connes time parameter $\tau$, are
\begin{align}
    p_B = \frac{\delta \textbf{L}}{\delta \dot{q}_B} &= \frac{a}{2}\bigg[2\dot{q}_B +(\beta_1 +\beta_2)\dot{q}_F \bigg]\\ 
    p_F = \frac{\delta \textbf{L}}{\delta \dot{q}_F} &= \frac{a}{2}\bigg[\dot{q}_B (\beta_1 +\beta_2)+\beta_1 \dot{q}_F \beta_2 + \beta_2 \dot{q}_F \beta_1 \bigg]
    \end{align}
    where we have denoted $a\equiv L_p^2/L^2c^2$.
The first of these equations can be written as an eigenvalue equation, which as explained in \cite{maithresh2019}, results from defining the modified Dirac operator:
\begin{equation}
\frac{1}{Lc}\;  \frac{dq}{d\tau}\sim D \equiv D_B + D_F ; \qquad D_B \equiv \frac{1}{Lc}\;  \frac{dq_B}{d\tau} ; \qquad   D_F \equiv \frac{\beta_1 + \beta_2}{2Lc} \frac{dq_F}{d\tau}
\end{equation} 
We note that it is a constant operator, and we can also express this as an eigenvalue equation
\begin{equation}
[D_B + D_F] \psi = \lambda \psi \equiv (\lambda_R + i \lambda_I)\psi \equiv \bigg(\frac{1}{L} + i \frac{1}{L_I}\bigg)\psi
\label{dirm}
\end{equation}
Since $D$ is bosonic, we assumed that the eigenvalues $\lambda$ are complex numbers, and separated each eigenvalue into its real and imaginary part. Furthermore, this is taken as the definition of the length scale $L$ introduced above. Moreover, as demonstrated in \cite{maithresh2019}, $L^3 = L_p^2 L_I$, and since $L$ is Compton wavelength, this implies $L_I = \hbar^2 /Gm^3$. 

These two lengths, $L$ and $L_I$, play a crucial role in what follows. $L$ is a measure of quantum dispersion, whereas the decoherence  length $L_I$ is a measure of spontaneous localisation. If $L<L_I$, quantum dispersion dominates classicality, and the STM atom behaves as a quantum elementary particle. If $L>L_I$, spontaneous localisation and classical behaviour dominates quantum dispersion and the STM atom behaves like a classical object. The quantum-to-classical transition takes place at $L=L_I$, in which case both the lengths are automatically equal to Planck length. This is the space-time-matter foam which exists at the Planck scale: it represents Planck scale quantum dispersion and spontaneous localisation, which is equivalent to a stochastic oscillation of Planck scale curvature [curvature is of the order 
$L^2 / L_p^4$]. The importance of Planck scale foam for the cosmological constant problem and for dark energy has been recently discussed also by \cite{Ng:2019,carlip2019,unruh2019}. In our theory, quantum gravity arises on time scales larger than Planck time, {\it after} coarse graining over Planck scale foam. The above relation between $L$ and $L_I$ can be more meaningfully written as
$(L/L_p)^3 = (L_I/L_p)$ and is very closely related to the Karolyhazy uncertainty relation \cite{SinghqgV2019}.
We note that if $L_I<L_p$, then both lengths are smaller than Planck length - this is the classical limit (it  corresponds to the associated Schwarzschild radius exceeding Compton wavelength). On the other hand, if $L_I > L_p$, then both lengths exceed Planck length, this being the quantum limit. The two lengths also define two important time scales: $t_q=L/c$ which is the quantum dispersion time scale, and $t_d=L_I/c$, the decoherence time scale. Classicality results if $t_d < t_q$, and quantum
behaviour prevails if $t_q < t_d$. 

These above are the defining equations for our dark energy particle. {We propose to name these particles as mitrons.} It is a quantum gravitational entity, with its gravitation aspect described by the bosonic $q_B$ and its matter aspect described by the fermionic $q_F$. The gravitational aspect of a dark energy particle cannot be described classically, because its decoherence length $L_I$ is enormous. For our assumed  mass of $10^{-33} \ {\rm eV/c^2}$, this length is $123$ orders of magnitude larger than the size of the observed universe. Consequently, the decoherence time $L_I/c$ is $10^{140}$ s, which of course is far far greater than the age of the universe. As a result the dark energy particle never undergoes the classicalisation process of spontaneous localisation. It is inherently quantum gravitational in nature. We also make the important assumption that dark energy particles do not entangle with each other, nor with other particles.

On the other hand, as explained in our earlier work \cite{maithresh2019, RMP:2012}, ordinary matter undergoes spontaneous localisation, because of entanglement, and gives rise to the observed classical universe. Against the backdrop of the classical expanding universe, let us now understand why the said dark energy particles are responsible for the acceleration of the universe. The contribution to the mass density of the universe, from a single dark energy particle, is of the order $m_{DE}/R_H^3$, where the mass $m_{DE}$ is of the order $10^{-66}$ gm, and $R_H$ the Hubble radius is about $10^{28}$ cm. This gives an extremely low mass density of the order of $10^{-150}$ gm/cc. 

We now recall from  recent discussion of the Karolyhazy uncertainty relation \cite{SinghqgV2019}, and from the implied holography, that the universe has $(R_H/L_p)^2\sim 10^{122}$  quanta of information \cite{Ng:2019}. Assuming one unit of information per elementary particle, we realise that ordinary matter contributes only a very small fraction to the information content, there being some $10^{80}$ such particles. Thus we propose, following Ng \cite{Ng:2019},  that there are $10^{122}$ dark energy particles in the observed universe, and hence their total contribution to the energy density of the universe is about $10^{-28}$ gm/cc, which is a reasonable ball-park estimate for the inferred dark energy content of the universe. To summarise, we are proposing that dark energy consists of $10^{122}$ quantum gravitational particles, each of mass $10^{-33}\ {\rm eV/c^2}$. Further, these STM atoms are assumed to be unentangled, and hence they do not undergo spontaneous localisation during the lifetime of the present universe. We might think of each such particle as a quantum gravitational wave of the size of the observed universe.

As noted by Ng, such objects collectively behave like dark energy, causing acceleration of the expanding universe, because they have a very long wavelength, and  their kinetic energy $T$ is negligible compared to their potential energy $V$. The pressure is essentially $T-V$ and the energy density is $T+V$. With $T$ being negligible, this implies an equation of state as for the cosmological constant: $p=-\rho$. In our theory, we can argue for the dominance of the gravitational aspect over the matter aspect by examining the expression for the Hamiltonian for the STM atom, given by \cite{maithresh2019}
\begin{equation}
   \textbf{H} = \text{Tr}\bigg[\frac{a}{2} (\dot{q}_B+\beta_1\dot{q}_F)(\dot{q}_B+\beta_2\dot{q}_F) \bigg]
\end{equation}
and in terms of the momenta
\begin{equation}
    \textbf{H} = \text{Tr} \frac{2}{a} \bigg[(p_B\beta_1-p_F)(\beta_2-\beta_1)^{-1}(p_B\beta_2-p_F)(\beta_1-\beta_2)^{-1}
    \bigg]
\end{equation}
From the first of these expressions it is evident that, since the $\beta$ matrices scale as 
$L_p^2/L^2$, and since $L\gg L_p$, the fermionic matter term is negligible, and the effective contribution to the Hamiltonian is from the gravitational term $\dot{q}_B^2$. In terms of the effective contribution of these dark energy STM atoms to the energy-momentum-tensor in Einstein's equations, this means that the kinetic energy term is negligible compared to the gravitational energy, and once again we see that the contribution of the dark energy particles behaves like a cosmological constant. The mass density estimate above shows that it is a cosmological constant of the same magnitude as implied by observations.

The above argument for neglecting the fermionic matter part, as compared to the bosonic gravitational part, does not apply to classical macroscopic objects, because they do not obey these quantum gravitational equations. Macroscopic objects obey classical Einstein equations and the accompanying matter equations of motion, and the relative magnitude of kinetic and potential energy terms is determined from the classical equations, in the conventional manner. (Note also that in the classical case $L\ll L_p$).  If we consider an ordinary quantum mechanical particle, such as a free electron or a proton, then again the gravitational aspect $\dot{q}_B$ will far exceed $\dot{q}_F$, because here too $L\gg L_p$. However, for such a particle, $L\ll R_H$, and we refer the quantum motion of the particle to the background space-time produced by the dominating classical bodies. The corresponding $\dot{q}_B$ of the quantum particle is not relevant, because it contributes negligibly to the classical space-time background. For the dark energy particles, their evolution cannot be referred to the background classical space-time, because their associated length $L$ is of the order of $R_H$. The quantum dark energy particles are not a perturbation to the space-time background produced by macroscopic bodies. Rather they dominate the classical background of the universe and dictate its expansion and acceleration.

Our theory seems to make a connection with Dirac's large number hypothesis. Integral powers of a large number in the range $\sim 10^{20} - 10^{21}$ occur frequently in our theory. Here we list various such large numbers which arise in our work. The ratio of the Compton wavelength of the proton to Planck length is about $10^{20}$. The related decoherence length $L_I = L^3/L_p^2 = \hbar^2/Gm^3$ is about $10^{27}$ cm. Its ratio to Planck length is $10^{60}$, this being the cube of $10^{20}$. The ratio of the Hubble radius to Compton wavelength of the proton is about $10^{41}$, and is close to the square of $10^{20}$. The ratio of Planck mass to proton mass is about $10^{19}$. The ratio of Planck mass to the mass of our newly introduced  dark energy particle is about $10^{61}$, being close to the third power of $10^{20}$. The number of particles of ordinary matter in the universe is about $10^{80}$. The number of information bits in our (holographic) universe, as also the number of dark energy particles, is about $10^{122}$, being close to the sixth power of $10^{20}$.
The spontaneous localisation lifetime of the proton (the so-called GRW value) is about $10^{17}$ sec, being $10^{60}$ in Planck units. The corresponding lifetime for the dark energy particle is $10^{183}$ in Planck units, being close to the ninth power of $10^{20}$.

The recurrent occurrence of powers of $10^{20}$ motivates us to tentatively suggest a dark matter candidate particle of mass about $10^{-12}\ {\rm eV/c^2}$, which is $10^{21}$ orders of magnitude lighter than the proton, and $10^{21}$ orders of magnitude heavier than the dark energy particle. We assume that the number of such dark matter particles in the observed universe is $10^{21}$ times more than the $10^{80}$ particles of ordinary matter. There are thus $10^{101}$ dark matter particles in the universe, and their mean density is $10^{-28}$ gm/cc. 
Assuming that there are about a billion galaxies in the observed universe, we could very roughly associate some $10^{91}$ dark matter particles per galactic halo on the average. This amounts to a dark matter mass of about $10^{46}$ gms per dark halo. The difference from the dark energy particle is that we assume the dark matter particles in a halo to be entangled, which results in rapid spontaneous localisation, allowing bound structures to form.
The classicalisation lifetime for a single dark matter particle  (i.e. the decoherence time) is $L_I/c=\hbar^2/Gm^3c$ which for the assumed mass is huge, some sixty-three orders of magnitude larger than the age of the universe. But if we assume that the $10^{91}$ particles in the galactic dark halo are entangled, this decoherence time comes down by $91\times 3$ orders of magnitude, and becomes much much less than one second. In fact to bring the decoherence life-time down to a millionth of a second, it is enough to entangle $10^{31}$ dark matter particles - the total mass of these many dark matter particles is nearly the same as that of a single proton. 

Does the dark energy in our model behave like the cosmological constant, or is it dynamical? We can argue as follows. The Compton wavelengths associated with a proton and a dark matter particle are $10^{-13}$ cm and $10^{8}$ cm respectively. Since these are both much smaller than the Hubble radius, they can be assumed to be decoupled from the expansion of the universe, and hence constant. On the other hand, the Compton wavelength for the dark energy particle is of the order of the Hubble radius, so we assume it to stretch in linear proportion to the Hubble radius. Thus the equivalent mass density of the dark energy particle falls as the inverse fourth power of the Hubble radius as the universe expands. Since the number of dark energy particles (the bits of information) increases as the square of the Hubble radius, the net dark energy density falls as the inverse square of the Hubble radius, and hence the dark energy is dynamical. Moreover, since it falls more slowly than ordinary matter density as well as slower than dark matter density, the universe was matter dominated in the past. Only in the present epoch does the universe become dark energy dominated. We emphasize that in our approach, dark energy is a purely quantum gravitational effect and can be properly described only in a quantum theory of gravity. Its description in the context of classical general relativity is only approximate. Such dark energy should be regarded as evidence for the quantum nature of gravity.

We note that our proposed particles, along with the proton, and Planck mass, form the following  mass scale (expressed in energy units): $10^{-33}$ eV [dark energy], $10^{-12}$ eV [dark matter], $10^{9}$ eV [ordinary matter], $10^{21}$ GeV [hundred times Planck mass]. Each  successive value is twenty-one orders higher than the previous one. We have already described quantum gravitational space-time-matter foam at the Planck mass/length scale. Masses above Planck mass are known to behave classically. If this pattern of masses is any indication, we should not expect to find any new particles heavier than the ones already known. This would suggest that beyond the standard model physics should be looked for at lower particle masses, such as the dark energy and dark matter particles proposed here, rather than at particle masses higher than the ones known.

In principle, there also exists the possibility that the dark matter particle is not independent of the dark energy 
particle.  {By this we mean that the dark matter particle is not elementary, but is a composite object made of $10^{21}$ mitrons. Such a composite state could be assumed to result from the quantum entanglement of $10^{21}$ mitrons.}
 The entanglement is equivalent to producing a bound state, in much the same spirit that the entanglement of say $10^{23}$ nucleons produces a macroscopic object, this being a bound state of enormously many ordinary atoms. If this were to be true, dark energy and dark matter would have their origin in the same ultra-light particle of mass $10^{-33}\ {\rm eV/c^2}$, and would represent different degrees of entanglement amongst many copies of this particle.

Lastly, we point out that the occurrence of such a dark energy can be motivated also from a duality property of the modified Dirac equation  (\ref{dirm}) which we reproduce below:
\begin{equation}
[D_B + D_F] \psi = \lambda \psi \equiv (\lambda_R + i \lambda_I)\psi \equiv \bigg(\frac{1}{L} + i \frac{L_P^2}{L^3}\bigg)\psi
\label{dirm2}
\end{equation}
which can also be written as
\begin{equation}
\left(\dot{q}_B  + i\beta \frac{L_P^2}{L^2}\dot{q}_F\right)\psi = \left(1+i\frac{L_P^2}{L^2}\right)\psi
\label{dirm3}
\end{equation}
where $\beta= (L^2/2L_P^2) (\beta_1 + \beta_2)$.  Let us define a new dual Compton length $L' = L_P^2 /L$, substitute it in this equation, and take its adjoint, to obtain
\begin{equation}
\left(\dot{q}_F^{\dagger} \beta^{\dagger}  + i \frac{{L}^2}{{L_P}^2}\dot{q}_B\right) \psi^{\dagger}= \left(1+i\frac{L_P^2}{{L'}^2}\right)\psi^{\dagger}
\end{equation}
We conclude that if $\psi$ is a solution for the STM atom $(q_ B, q_F)$ with Compton wavelength $L$, then $\psi^{\dagger}$ is a solution for the dual atom $(q_B', q_F')$ with Compton wavelength $L'$ where the primed variables are related to the unprimed ones by the relation
\begin{equation}
\dot{q}_F^{\dagger} \beta^{\dagger}  + i \frac{{L}^2}{{L_P}^2}\dot{q}_B = \dot{q'}_B  + i\beta \frac{{L_P}^2}{{L'}^2}\dot{q'}_F
\end{equation}
so that we get
\begin{equation}
\left(\dot{q'}_B  + i\beta \frac{L_P^2}{{L'}^2}\dot{q'}_F\right)\psi^{\dagger} = \left(1+i\frac{L_P^2}{{L'}^2}\right)\psi^{\dagger}
\end{equation}
in complete correspondence with (\ref{dirm3}). An STM atom with length $L$ greater than Planck length (equivalently mass $m$ less than Planck mass) is dual to another STM atom with dual length $L'$ less than Planck length (equivalently dual mass $m'=m_{Pl}^2 / m$ greater than Planck mass). For our dark energy particle with mass $10^{-66}$ gm, the dual object has a mass $10^{56}$ gm, which happens to be the mass of the observed universe. This is an intriguing relation between the dark energy particle and the observed universe.

The modified Dirac equation also has important implications for how the classical limit emerges or is avoided.  A classical object has Compton wavelength $L$ much less than Planck length, and as expected, the imaginary part of the eigenvalue dominates over the real part, in Eqn. (9). Such an object, having a size $L< L_P$, experiences the ever-present Planck scale foam in its dynamics. The foam, having random oscillations in curvature \cite{SinghqgV2019}, decoheres the quantum motion, rendering the object classical. We recall that the statistical thermodynamics of the underlying matrix dynamics of STM atoms requires averaging over time scales much larger than Planck time. This is allowed for those STM atoms for which the length  $L$ is much larger than Planck scale. But if $L$ is smaller than Planck length, the condition for the validity of equilibrium statistical mechanics break down. The STM atom is subject to extremely rapid Planck scale fluctuations, which render it classical. Thus it is evident that classical behaviour is arising because of dynamics taking place at the Planck scale. In other words, Planck scale quantum gravity is responsible for the resolution of the quantum measurement problem. On the other extreme, the dark energy particle has a size $L$ far far greater than Planck length, and is not sensitive to Planck length physics, and remains entirely quantum.

\vskip 0.4 in

\centerline{\bf  {DISCUSSION}}

The following discussion clarifies further various aspects of our proposal, and puts it in the context of other similar approaches to 
quantum gravity and dark energy.

\begin{itemize}

\item {\it Understanding the dark energy degrees of freedom}: It is perhaps not quite appropriate to think of the dark energy degrees of freedom in quantum field theoretic concepts. An STM atom is neither fermionic nor bosonic, but is rather a combination of both aspects. It is reasonable to think of these particles as indistinguishable, but without attributing a bosonic character or a symmetrized state. In fact the dynamics is not given by quantum field theory, but by a [non-unitary] matrix dynamics. Its closest analog would be the Heisenberg picture (though the two are not identical) - for a comparison of the matrix dynamics with the Heisenberg picture, see e.g. the discussion in \cite{Adler:04}. With regard to the associated statistics, perhaps a good possibility is the so-called infinite statistics, discussed by \cite{Ng} in a similar context as here, where he also considers dark energy to be extremely light Hubble scale particles.

\item {\it Dark energy interactions}: One could well ask the following question, in the context of the assumed properties of the dark energy particles: Why is no interaction of the dark energy particle allowed with each other or any other degree of freedom? This seems a rather ad-hoc assumption and in particular seems to be in violation of the universal coupling of gravity with energy-momentum.

To answer this, we must first distinguish between gravity and other interactions. At the level of the Planck scale matrix dynamics, each STM atom has a gravitational aspect, coded in $q_B$. However, this gravity is not an interaction between STM atoms. The notion of a gravitation interaction a priori assumes the existence of a background space-time, which is not there in the Planck scale matrix dynamics. When a large collection of entangled STM atoms undergo spontaneous localisation, classical space-time and classical gravitation emerge, and these obey Einstein equations, with the fermionic (matter) part acting as a 'source' for the bosonic (gravity) part.  

On the other hand, even in the Planck scale matrix dynamics, it is possible to talk of Yang-Mills gauge interactions between STM atoms \cite{Abhishek2020}, without the need for a background spacetime. For those STM atoms which have a Yang-Mills interaction, entanglement is possible, and this entanglement is crucial in enforcing spontaneous localisation and space-time emergence. It is an assumption of our model, to be investigated in further research, that the dark energy particles do not possess Yang-Mills interactions. However, as noted above, they do have a gravitational aspect each, though it not meaningful to talk of gravitational interaction {\it between} them. Classical gravity is an emergent concept.

\item{\it Comparison with the cosmological constant}: The accelerated expansion of the universe is currently attributed to a cosmological constant, which contributes an energy density of about $(2\times 10^{-3}\ eV)^4$. From a conventional viewpoint of how quantum vacuum energy might contribute to gravity, the expected value of the constant is on the Planck scale: $(10^{19}\ GeV)^4$, which is about $10^{122}$ orders of magnitude higher. This of course is the notorious cosmological constant problem. Another way to express the problem is that the observed cosmological constant is of the order $1/R^2\sim 10^{-56}$ cm$^{-2}$ 
where $R\sim 10^{28}$ cm is the size of the observed universe. This is a factor $(L_p/R)^2 \sim 10^{122}$  smaller than the natural value of $L_p^{-2}$. However, this fine-tuning problem does not arise in our theory, for the following reason. In the conventional thinking, this quantum vacuum energy should act as a source for gravity, on the right hand side of Einstein equations. However, this is not the correct way to couple a quantum system to gravity. Quantum systems exhibit superpositions, and do not produce classical gravitational fields. In our approach, we have proposed that the gravity of a quantum system must be described by the matrix dynamics of STM atoms, as explained above. This is a very different description from Einstein equations, with the latter coming into play only in the classical limit, after space-time emergence.

The observed dark energy density of $(2\times 10^{-3}\ eV)^4$ is divided, in our theory, amongst $(L_p/R)^2 \sim 10^{122}$ dark energy particles, the so-called mitrons, for holographic reasons described above in the paper. This is what gives each mitron the assumed mass of about $10^{-33}$ eV/c$^{2}$. Thus the assumed value of the mass is a direct consequence of requirement that it produce the observed dark energy density. A given mitron {\it does not} produce a classical gravitational field equivalent to that of a classical particle with this much mass. This object, having a associated length of the order of Hubble radius, is not embedded in a  classical (Robertson-Walker) space-time, and hence is not a source for its geometry in the Einstein sense. Only effectively, the gravitation of all the mitrons put together acts like dark energy, as demonstrated above.

By choosing by hand the desired value of the mass of the mitron, have we exchanged one fine-tuning problem for another. No, in a sense. Because holography, along with the observed cosmic acceleration,  strongly suggests that such particles must exist. The existence of an ultra-light particle of mass $10^{-33}$ eV/c$^{2}$ becomes a prediction of our theory, and poses a challenge for experimental particle physics to detect it, directly or indirectly.

As for a geometric origin of the cosmological constant in Einstein equations, we note that in the heat kernel expansion of the square of the Dirac operator $D_B$, there does arise a cosmological constant term, but at order $L_p^{-4}$. There is no such term at order $L_p^0$ or at order $l_p^2$. Thus, there is no freedom left to add such a term by hand, when one treats general relativity as an emergent theory. As for the term at order $1/L_p^4$, its interpretation is not clear to us at present, and this issue is left for future investigation.

\item{\it Comparison with other approaches}: The theory closest to our work is trace dynamics, which is a matrix dynamics of fermions and gauge fields, at the Planck scale. Here too, the attempt is to derive quantum field theory as a low energy approximation to the matrix dynamics. However, since gravity is not included here, the issue of dark energy and cosmological constant is not directly addressed. 

There have been other interesting `quantum-first' approaches to gravity, for example the work of Giddings \cite{Giddings1, g2} and Carroll and collaborators \cite{Carroll1, c2}. The idea here is that instead of quantising an already given classical theory of gravity, one looks to add fundamental structure to quantum mechanics, which would enable the inclusion of gravity [in a quantum gravity sense], and from which classical space-time geometry will be emergent, possibly as a consequence of entanglement. There are important commonalties between these approaches, and ours. The common goal is that something should be done to quantum mechanics so as to include gravity in it, and also to make key use of entanglement. What we `do' to quantum theory is to remove classical space-time from it. However, there are important differences too, from these approaches. These approaches would like to retain the concepts of unitarity and locality/separability. The matrix dynamics we construct is non-unitary, with unitary quantum field recovered as a low energy approximation, in the limit of sub-critical entanglement. Also, the matrix dynamics is separable in the sense that the different STM atoms are enumerable, but the dynamics is not local, in the sense that space-time and matter are not distinct from each other. Classical space-time, locality, and material separability are recovered in a low energy approximation, as a consequence of super-critical entanglement. The presence of an anti-self-adjoint part in the matrix Hamiltonian assists the entangled STM atoms to undergo dynamical spontaneous localisation, giving rise to emergent locality, separability, and classicality. This same non-unitary aspects permits the Karolyhazy relation to arise, and hence also the quantum gravitational dark energy. To our understanding, in the other quantum-first gravity approaches, the status of the cosmological constant and vacuum energy  is not changed.

\item{\it Dirac's large number hypothesis}: Although powers of $10^{20}$ occur widely in our work, and in related situations, we have not found a theoretical proof as to why this should be so. To that extent their occurrence could well be a coincidence. Except to note that a key role might be played by the ratio $L^2/L_p^2 \sim 10^{40}$, where $L$ is the Compton wavelength of the proton. Why is this Compton wavelength so much smaller than Planck length, and does it have anything to do with the fact that the size of the observed universe is $10^{60}$ times larger than Planck length? We also note from our relation
$L^3 = L_p^2\; L_I$ that if $L_I$ is taken to be the size of the observed universe, then $L$ is very nearly the Compton wavelength of the proton.

Our assumed value for the mass of the dark matter particle ($10^{-12}$) eV, crucially depends on the Dirac hypothesis being not a coincidence, but of some fundamental origin. Furthermore, if it is indeed true that these particles are not elementary, but composite of dark energy particles resulting from interaction and entanglement, there would have to be some constraint at the initial big bang event, because of which some of the dark energy particles were entangled, whereas others were not. This remains an unresolved aspect for future investigation.

\item{\it Holography and the bulk-boundary correspondence}: In our theory, holography is justified because for a region of size $L_I$ with cells of size $L$, we have because of the Karolyhazy relation that $L_I^3 / L^3 = L_I^2 / L^2$. Implying that the amount of information in the region grows as the area of its boundary. This would suggest that the d.o.f. associated with the dark energy STM atoms `reside' in the two-dimensional boundary of the region, in a sense at the spatial boundary of the observed universe. Intriguingly, STM atoms can be said to be 2-d objects in the matrix dynamics. This is evident from the structure of the fundamental action (\ref{funacn}). The two $\beta$ matrices being unequal, gives this action a two-dimensional structure, as opposed to the 1-d structure suggested by the configuration variable $q$. [The comparison with the 2-d string of string theory is suggestive!]. Thus it remains to be understood if the 2-d DE particles bear any relationship to the spatial boundary of the observed universe. At the level of matrix dynamics, the DE particles evolve in Connes time, as is evident from the action principle, making them 2+1 dimensional entities. There is a bulk-boundary correspondence: the observed universe is the bulk. The DE particles in the matrix dynamics are the boundary. We have already seen above, from the modified Dirac equation,  a duality between the DE particles and the observed universe.

It would also be worth exploring in the future if there is a connection with the highly successful AdS/CFT correspondence, or a possible similar conjecture for the de Sitter case. For, after all, the DE particles which are dominating the current expansion are the ones which lend the deSitter structure to which the present universe is asymptotically approaching. It would be important to explore if the 2-d matrix dynamics of the STM atoms has something to do with a CFT. Also, we have earlier shown that \cite{maithresh2019b} the Bekenstein-Hawking entropy for a black hole can be explained as originating from the spontaneous localisation of a large collection of entangled STM atoms. In an analogous manner, it will be interesting to investigate if the  entropy  associated with the de Sitter universe can also be derived as a spontaneous localisation process, starting from the underlying matrix dynamics.

\item{\it Schwarzschild energy bound of the universe}: The duality pointed out towards the end of the paper, between a mitron and the entire universe, is also an instance of particle - black hole duality. A particle of mass $m_F$ much smaller than Planck mass  is dual to a black hole of mass $m_{BH}=m_p^2/m_F$, in the sense that the two solutions can be mapped to each other \cite{maithresh2019}. Thus, when the particle is a mitron, its dual is a black hole with the same mass as the entire universe. In fact the Compton length of the mitron is equal to the Schwarzschild radius of a black hole with the same mass as the entire universe! This suggests a deep connection between the properties of a mitron and the observed universe considered in its entirety.

\end{itemize}

\vskip 0.4 in

\centerline{\bf {APPENDIX}} 

\section{A BRIEF REVIEW OF THE NEW PLANCK SCALE MATRIX DYNAMICS}

Starting from a general classical dynamics, the canonical degrees of freedom [i.e. the configuration variables and momenta] are raised to the status of matrices (equivalently, operators). This is the same first step that is taken when we (canonically) quantise a classical system; except that we do not impose quantum commutation relations. The commutation relations between all dynamical variables are now determined by the initial conditions and the dynamics, as follows. The original Lagrangian of the classical theory now becomes an operator polynomial; a matrix trace is constructed from it, and this scalar serves as the Lagrangian of the new theory, referred to as the trace Lagrangian. A Lagrangian dynamics is constructed from here, by extremising this Lagrangian with respect to the configuration matrices. These matrices themselves are assumed to be made of Grassmann elements. There are two kinds of matrices: those made of odd-grade Grassmann elements (fermionic matrices) and those made of even-grade Grassmann elements (bosonic matrices). They respectively describe fermionic and bosonic degrees of freedom, as in quantum field theory.

The above is the essence of Adler's theory of trace dynamics, assumed to operate at the Planck scale. However, space-time is assumed to be Minkowski, and gravity has not been included in this theory. From this matrix dynamics, quantum field theory is derived as a low energy emergent approximation, by using the techniques of conventional statistical thermodynamics. The idea being that the original matrix dynamics is not being examined at Planck time resolution [equivalently Planck energies] but at lower energies. Hence a coarse-graining over many Planck time intervals is carried out, to find out the mean dynamics.

In our theory, we have proposed how to include gravity in Adler's theory of trace dynamics, by appealing to Connes' 
non-commutative geometry (NCG) programme. According to a result in geometry, it is possible to represent the curvature information of a Riemannian space-time in a spectral manner, through the spectrum of the conventional Dirac operator on the space-time. A heat kernel expansion of the Dirac operator $D_B$, truncated at the order $L_p^{-2}$, in an expansion in powers of square of Planck length, shows that
\begin{equation}
Tr [L_P^2 \ D_B^2] \propto L_P^{-2} \ \int d^4x \ \sqrt{g} \ R
\end{equation}
where $R$ is the Ricci scalar. The eigenvalues of the Dirac operator capture the information about curvature in a spectral way, and this opens the way for spectrally capturing geometric information when the manifold is replaced by a non-commutative space. Say, as in our case, by raising the coordinates on the manifold to the status of (non-commuting) matrices. Even though the 
space-time manifold is lost, and so is the concept of Riemannian curvature, the Dirac operator continues to exist, and is assumed, through its eigenvalues, to describe `curvature' of the non-commutative space. This is the transition to non-commutative geometry. Furthermore, and very importantly so, the non-commutative algebra associated with this non-commutative geometry admits a one-parameter family of automorphisms of the algebra, and this scalar parameter serves to play the role of time [what we have denoted as Connes time $\tau$]. This is a consequence of the Tomita-Takesaki theory, and is a property unique to 
non-commutative geometries - there is no such analog in the commutative case. Thus although the concept of space-time is lost, a valuable notion of time is recovered, which allows us to define dynamics at the Planck scale.

In our work, considering that the Dirac operator has dimensions of inverse length, we defined a bosonic `gravitational' configuration operator $q_B$ by the relation $D_B \equiv (1/L) \ dq_B/d\tau$. We come back to significance of this length $L$ shortly. Since we want to include gravity in trace dynamics, we ask: how might one describe the gravitational effect of a quantum particle such as an electron? Since the particle and its gravitational effect are both delocalised, we do not make a distinction between the two; hence the concept of an atom of space-time-matter. An STM electron describes both the electron and the `gravitation' it produces. We denote this object by an operator $q$ made of Grassmann elements, and we write $q$ as a sum of a bosonic matrix and a fermionic matrix (this can always be done): $q=q_B + q_F$, and we may think of $q_B$ and $q_F$ as respectively the gravity (bosonic) and the matter (fermionic) aspect of the STM electron. The fermionic part $D_F$ of the gereralised Dirac operator $D\sim  D_B + D_F$ is defined as $D_F\equiv dq_F/d\tau$. of the The action principle for an STM atom is proposed to be, as described earlier in this paper:
 \begin{equation}
\frac{L_p}{c} \frac{S}{C_0}  =  \frac{1}{2} \int d\tau \; Tr \bigg[\frac{L_p^2}{L^2 c^2}\; \left(\dot{q}_B + \frac{L_P^2}{L^2}\beta_{11} \; \dot{q}_F\right)\;\left(\dot{q}_B +\frac{L_P^2}{L^2}\beta_{22} \; \dot{q}_F\right) \bigg]
\end{equation}
There is one such term for each STM atom, and the total action is additive. We explicitly assume that the two $\beta$ matrices are each proportional to $L_p^2 / L^2$. Such proportionality can be naturally expected from a heat-kernel expansion of the Dirac operator. Moreover, this proportionality is absolutely essential for recovering the correct matter-gravity coupling when the Einstein equations of general relativity are recovered in the classical limit. 

The length scale $L$ associated with an STM atom is its only parameter; one does not associate a mass or spin with it in the Planck scale dynamics. Mass and spin are only low-energy emergent concepts. Thus an STM atom is neither bosonic nor fermionic: the Grassmann elements of the matrix describing it are not restricted to be odd-grade or even-grade. Moreover, one could well ask what is the interpretation of a length $L$ when there is no space-time? To answer this, we note that every physicsal quantity in the above action is scaled so as to be dimensionless. The action $S$ is scaled by a quantity $C_0$ with the dimensions of action; and the configuration variable $q$ having dimension of length is scaled by $c\tau$, where the speed of light $c$ is to be understood as the ratio $\\L_p/\tau_p$. The time $\tau_p$ is Planck time, which scales Connes time $\tau$. Thus the length $L$, which is scaled by Planck length, should be understood as the ratio $L/l_p$, and this is a dimensionless quantity (a number) associated with the STM atom. The length interpretation, strictly speaking, arises only after space-time emergence.

The generalised Dirac operator
\begin{equation}
\frac{1}{Lc}\;  \frac{dq}{d\tau}\sim D \equiv D_B + D_F ; \qquad D_B \equiv \frac{1}{Lc}\;  \frac{dq_B}{d\tau} ; \qquad   D_F \equiv \frac{\beta_1 + \beta_2}{2Lc} \frac{dq_F}{d\tau}
\end{equation} 
is bosonic but not self-adjoint, because while $q_B$ is self-adjoint, $q_F$ is not. Here, $\beta_1 = \frac{L_P^2}{L^2}\beta_{11}$ and $\beta_2 = \frac{L_P^2}{L^2}\beta_{22}$.
We note that it is a constant operator, and we can also express this as an eigenvalue equation
\begin{equation}
[D_B + D_F] \psi = \lambda \psi \equiv (\lambda_R + i \lambda_I)\psi \equiv \bigg(\frac{1}{L} + i \frac{1}{L_I}\bigg)\psi
\label{dirm}
\end{equation}
Since $D$ is bosonic, we assumed that the eigenvalues $\lambda$ are complex numbers, and separated each eigenvalue into its real and imaginary part. Furthermore, this is taken as the definition of the length scale $L$ introduced above. Moreover, as demonstrated in \cite{maithresh2019,SinghqgV2019}, $L^3 = L_p^2 L_I$, and since $L$ is Compton wavelength, this implies $L_I = \hbar^2 /Gm^3$. A direct way to understand this important relation between $L$ and $L_I$ is to note that the $\beta$ matrices are proportional to $L_p^2 /L^2$. It is precisely this proportionality (which was essential for recovering Einstein equations) which imposes this Karolyhazy uncertainty relation on our theory. Thus, we start by assuming that the real and imaginary parts of the eigenvalue are both proportional to $1/L$; however, because of the presence of the additional factor in $\beta$, the imaginary part is further suppressed by a factor $L_p^2 /L^2$. Thus the  real and imaginary parts are in the ratio
$1/L\ : \ (L_p^2/L^2)\; (1/L)$ - this explains the above relation. between $L$ and $L_I$. 

We also remark on the Connes time $\tau$ which results from the Tomita-Takesaki theory applied to our non-commutative matrix dynamics. The choice of $\tau$ requires a specification of the overall state for the matrix dynamics. We propose this state to be the one in which the universe was initially created as an enormous collection of STM atoms. While a proper understanding of this state requires further research, we have previously speculated that the so-called Big Bang event is a huge spontaneous localisation event, and the subsequent expansion of the universe can be thought of as the reverse of spontaneous localisation.

The emergence of classical space-time from our Planck scale matrix dynamics is explained in detail in our earlier papers
\cite{maithresh2019,Singh:sqg}. The essence of the emergence is that the Hamiltonian of an STM atom in the matrix dynamics is not self-adjoint. It has a tiny anti-self-adjoint part of the order $Lp/L$ which is negligible for elementary particles, since for them $L\gg L_p$. When we perform a statistical thermodynamics to find out the low-energy mean dynamics, there are two limiting cases. So long as not too many STM atoms are entangled with each other, the emergent theory is quantum field theory. However, if a sufficiently large number of STM atoms entangle, the effective length $L_{eff}$ associated with the entangled system goes below Planck length, and the anti-self-adjoint part of the total trace Hamiltonian can no longer be neglected. This contributes significantly to non-unitary evolution, resulting in spontaneous localisation of the fermionic part of the STM atoms, and thereby the emergence of classical space-time and its gravitation. Those STM atoms which are not significantly entangled behave like the quantum systems we know of, and obey the laws of quantum field theory.

The entanglement of STM atoms is very likely a consequence of their Yang-Mills interactions, which cause elementary particles to bind to each other. In this paper we have proposed that the dark-energy particles (dubbed the mitrons) are extremely light, with length $L$ of the order of Hubble radius, and not subject to the Yang-Mills interactions. Hence they do not entangle with each other, but they do have a gravitational aspect associated with them (the bosonic part of a mitron). But they cannot be described by the rules of quantum field theory, because they do not exist on a classical background space-time. Their dynamics can only be correctly described as the underlying matrix dynamics of STM atoms, in which there is no space-time.

\bigskip

\bigskip

\centerline{\bf REFERENCES}

\bibliography{biblioqmtstorsion}

\newpage

\end{document}